# Non-Linear pricing with differential machine learning

Pavel Goldin
Pavel_Goldin@epam.com

## Abstract

The objective of this research was to evaluate and gain experience with application of two methods used for pricing and sensitivity analysis of exotic financial derivative instruments, namely, automatic adjoint differentiation (AAD) and deep learning.

The work was inspired by publication [4] of Danske Bank quantitative analysts Antoine Savine and Brian Huge in which the authors introduced a novel approach to building extremely efficient pricing and risk approximators for arbitrary financial derivative instruments.

## Introduction

Differential machine learning (ML) presented in [4], combines automatic adjoint differentiation (AAD) [3] with deep learning to estimate value and risk sensitivities of the financial derivatives. Differential ML is a kind of supervised learning, where the models are trained on datasets (inputs & labels) *augmented with differentials of labels wrt inputs*. In the context of financial Derivatives and risk management, pathwise [3] differentials are efficiently computed with automatic adjoint differentiation (AAD). The pathwise estimator is calculated by interchanging, if it is possible, the order of differentiation and integration.

$$\frac{\partial}{\partial \theta} E[Y(\theta)] = E\left[\frac{\partial Y(\theta)}{\partial \theta}\right] \quad (1)$$

where $Y(\theta)$ is payoff.

For example, in the simplest situation in the Black-Scholes framework. While an explicit expression for the option delta is available, we can also estimate it via the pathwise method as follows. We first write the option payoff as

$$Y = e^{-rT}(S_T - K)^+ \quad (2)$$

$$S_T = S_0 e^{\left(r - \frac{\sigma^2}{2}\right)T + \sigma\sqrt{T}Z} \quad (3)$$

where $Z \sim N(0,1)$. It follows from (2) and (3) that

$$\frac{\partial Y}{\partial S_0} = \frac{\partial Y}{\partial S_T}\frac{\partial S_T}{\partial S_0} = e^{-rT} 1(S_T > K) \frac{S_T}{S_0} \quad (4)$$

The estimator (4) is easily calculated via a Monte-Carlo simulation.

The foundation of differential ML is twin network. Twin network combines two networks into a single representation, corresponding to the computation of a prediction (approximate price) together with its differentials wrt inputs (approximate risk sensitivities). The first part of the twin network predicts a value. The second part predicts risk sensitivities. It is the mirror image of the first half, with shared connection weights [4].

The strength of the approach is fast training. When learning Derivatives pricing and risk approximation, the main computation load belongs to the simulation of the training set. For complex products prices are computed numerically, generally by Monte-Carlo. Monte-Carlo valuation has highly unrealistic cost in a practical context. In this approach, sample datasets are produced for the computation cost of one Monte-Carlo pricing, where each example is one sample of the payoff, simulated for the cost of one Monte-Carlo path.

As a result, learning time is reduced dramatically and training sets are simulated in realistic time. Other strengths are accurate pricing of option values and 'greeks', online, in real time. The calculation speed of 'greeks' is not much more than speed of calculation of closed form solution. This methodology is applicable to arbitrary Derivatives instruments under arbitrary stochastic models of the underlying market variables. Differential machine learning, combined with ADD, provides extremely effective pricing and risk approximations. We can produce fast pricing analytics in models too complex for closed form solutions, extract the risk factors. Differential machine learning can speed up the performance of in-house models very strongly.

Also, this methodology can be used in high dimensional cases for large portfolios. To reduce dimensions helps PCA and differential PCA to learn NN more effectively, it helps to minimize the size of neutral network and the number of inputs. This approach can be also useful for execution algos because we can easily calculate 'greeks' and duration for different complex products.

# Cases selected to evaluate the approach of differential neural networks and estimate performance of pure AAD

1. Valuation of Asian options with arithmetic averaging and their greeks (delta and vega) for the model with constant volatility.
2. Valuation of Asian options with arithmetic averaging and their greeks (delta and vega) with arbitrary volatility curve.
3. Valuation of option written on the basket of correlated stocks. We also estimated gamma, option 2nd derivative.
4. Estimation of callable bond and its duration.
5. Differential PCA [4] for large portfolio of correlated instruments (250, 500 and 1000).
6. Libor Market Model [3]. Pricing options which payoffs depend on LIBOR rates.
7. Pricing of option and its vega for the model of stochastic volatility (SABR).
8. Valuation worst-of options and their greeks for the basket of n-correlated instruments,

1.  Valuation of Asian options and 'greeks' (delta and vega). Asian options are of particular importance for commodity products which have low trading volumes. The terminal payoff depends on some form of averaging of the underlying asset price over a part of or the whole of the option's life. When the option payoff depends on the average of the underlying asset over a time interval, the option tends to be less expensive than its European counterpart. Furthermore, the averaging feature can lessen incentives for market manipulation, and the volatility of an average is lower than the volatility of the underlying asset, thus explaining their usage in risk management. It makes Asian options ideal for use in hedging positions. We consider the case discrete arithmetic Asian call option with payoff

$$P(S) = Max\left(\frac{1}{m+1}\sum_{i=0}^{m} S\left(\frac{iT}{m}\right) - K, 0\right) \quad (5)$$

There are no known closed form analytical solutions to arithmetic average Asian options, many numerical methods are applied like Monte Carlo simulation, partial differential equations and moment matching method. In this experiment, we assume volatility as a constant parameter. It is rather simply to construct a pathwise estimator for the delta of this option [7].

$$\frac{\partial P(S)}{\partial S_0} = e^{-rT} 1(\check{S} > K) \frac{\check{S}}{S_0}, \quad \check{S} = \frac{1}{m+1}\sum_{i=0}^{m} S\left(\frac{iT}{m}\right) - K \quad (6)$$

We can get the pathwise estimator of the vega of the Asian option

$$\frac{\partial P(S)}{\partial \sigma} = e^{-rT} \mathbf{1}(\check{S} > K) \frac{1}{m+1} \sum_{i=0}^{m} \frac{\partial S\left(\frac{iT}{m}\right)}{\partial \sigma} \qquad (7)$$

It is easy to see that differential neutral network gives better accuracy than feed forward net especially for vega.

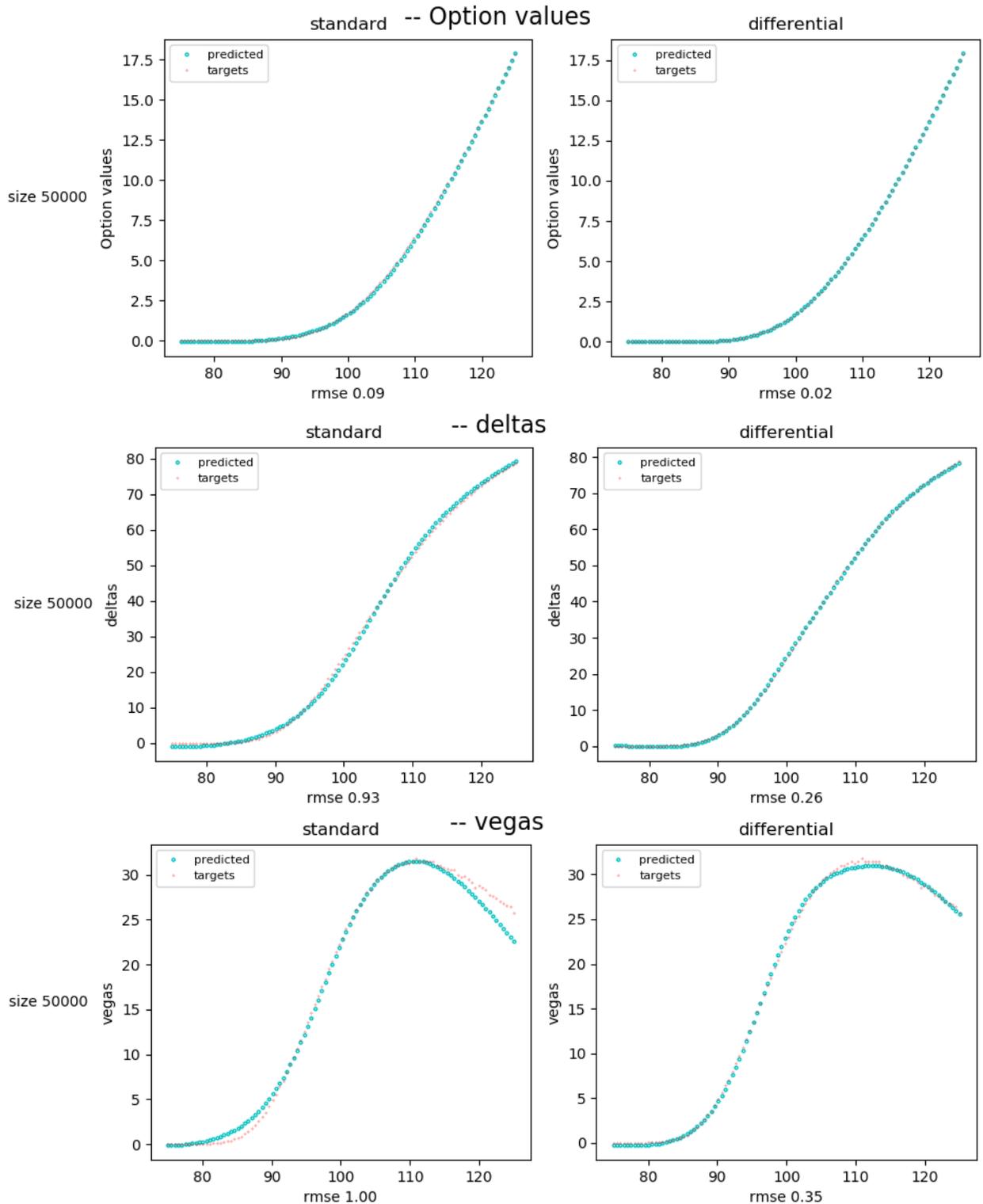

2. Valuation of Asian options and 'greeks'. In this approach we modeled the volatility by means volatility curve. It is very important for practitioners. We admit different volatilities for different averaging periods. Volatility curve can be extracted from option chain. In this situation we have sensitivity of option value for each volatility. It was found that the greater number of volatilities, the stronger advantage of differential neutral network. Differential neutral network outperforms feed forward net in high dimensional tasks.

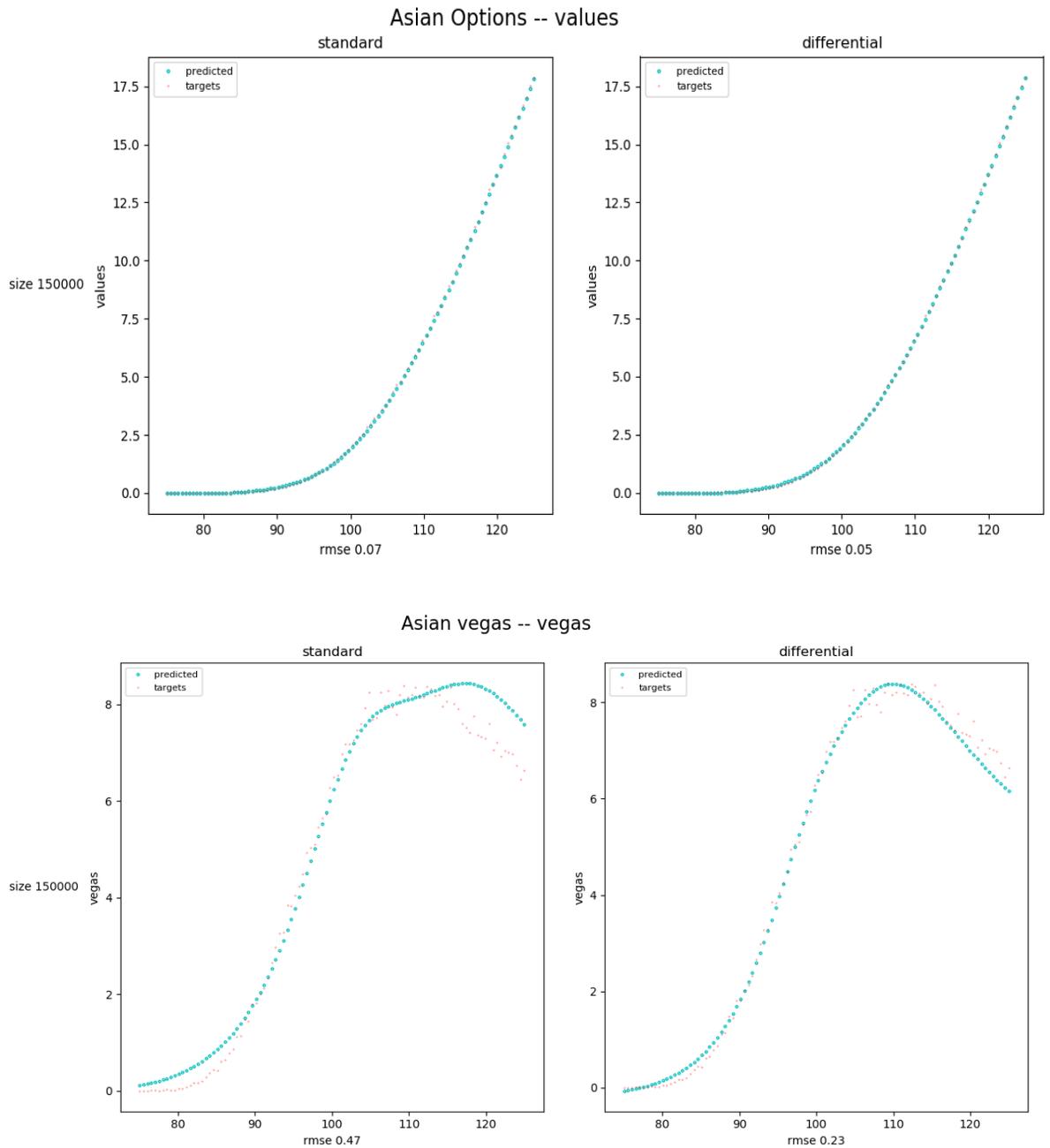

3. Valuation of option written on n correlated stocks. We use Multivariate log-normal distribution. We use method to generate correlation matrix in multivariate log-normal distribution with method suggested in [8]. This method produces a positive-semidefinite matrix, it is fast to implement even for large matrices, it allows the determination of a feasible matrix that most closely approximates a target real symmetric (but not positive-semidefinite) matrix. This method proposes for the construction of a valid correlation matrix $C = BB^T$ is to view the elements of the row vectors of matrix $B$ as coordinates lying on unit hypersphere. If we denote by $b_{ij}$ the elements of the matrix $B$, the key is to obtain the $n \times n$ coordinates $b_{ij}$ from $n \times (n-1)$ angular coordinates $\theta_{ij}$ according to

$$b_{ij} = \cos\theta_{ij} \prod_{k=1}^{j-1} \sin\theta_{ij} \quad for\ j = 1..n-1 \tag{8}$$

and

$$b_{ij} = \prod_{k=1}^{j-1} \sin\theta_{ij} \quad for\ j = n \tag{9}$$

Thanks to the trigonometric relationship and to the requirement that the radius of the unit hypersphere should be equal to one, the main diagonal elements are guaranteed to be unity. We generate price process by means follow stochastic differential equation:

$$S^i_{t+\Delta t} = S^i_t \exp\left(\left(\mu_i - \frac{\sigma_i^2}{2}\right)\Delta t + \alpha_i \sigma_i \sqrt{\Delta t}\right) \tag{10}$$

Where: $S^i_t$ denotes the price $S^i$ at time t and $(\alpha_1,..., \alpha_N)$ are derived by taking the Cholesky decomposition $LL^T$ of the "correlation matrix" and the applying it to N iid standard normal variables $(\epsilon_1,..., \epsilon_N)$.

We estimated gamma, it is second derivative. Gamma is used in option trading, it is important fast speed and high accuracy for this 'greek'. We combine the pathwise and likelihood ratio method for this objective. In contrast to the pathwise method, the likelihood ratio method differentiates a probability density with respect to the parameter of interest, $\theta$. It provides a good potential alternative to the pathwise method when payoff is not continuous in $\theta$ [7]. We rewrite

$$1(x > K) = fe(x) + \big(1(x > K) - fe(x)\big) \tag{11}$$

where $fe(x) = \min\left(1, \frac{\max(0, x-K+eps)}{2eps}\right)$ and $he(x) = 1(x > K) - fe(x)$. Here $fe(x)$ is piecewise linear approximation to the function 1(x>K) and that he(x) corrects the approximation, We applied the pathwise estimator to $fe(x)$ and likelihood estimator to $he(x)$. We provide result for 5-dimensional case. It is easy to see that accuracy of differential neutral network is better than accuracy of feed forwarf neutral network.

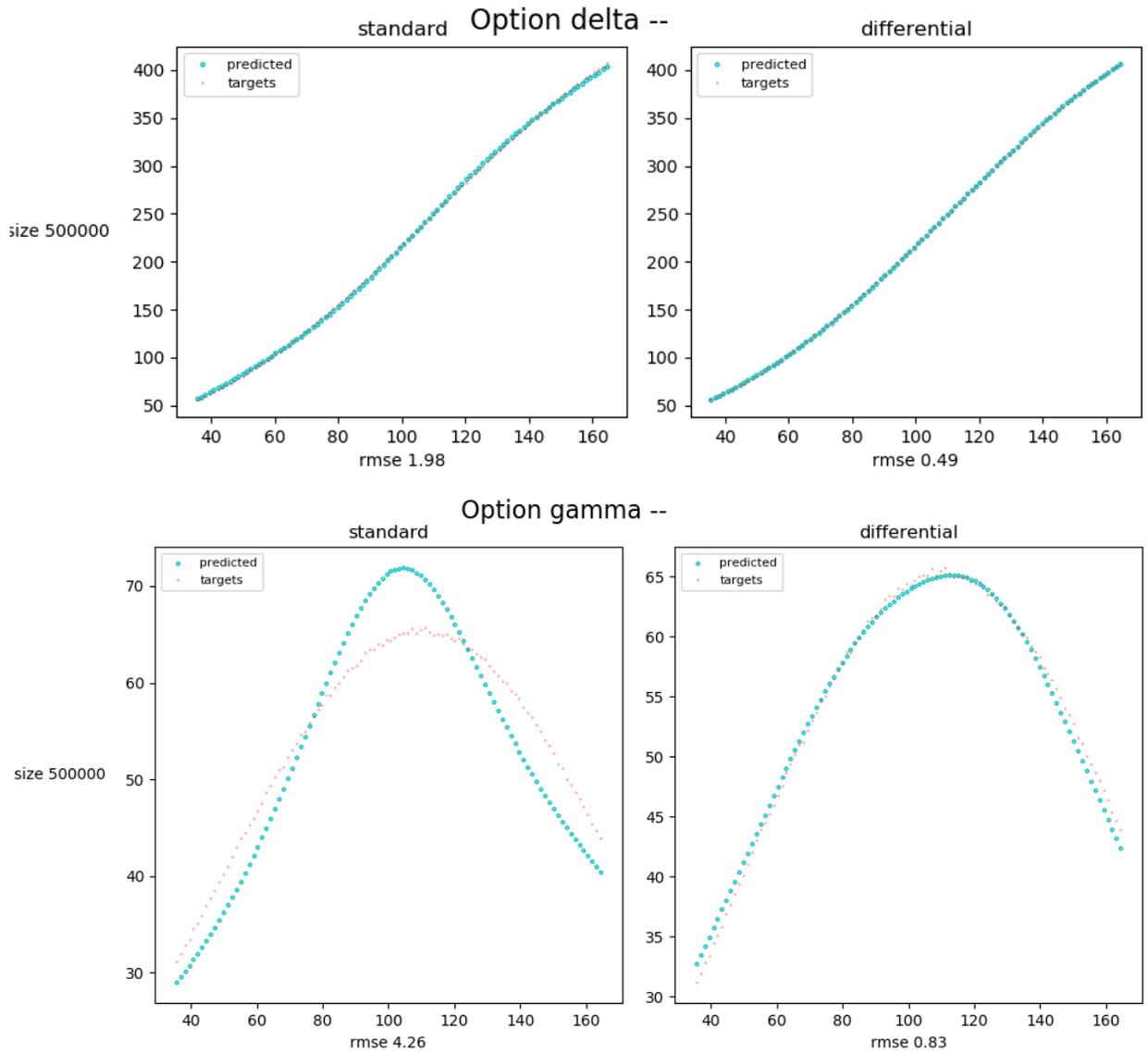

4. We estimated callable bond and duration. Fast estimation of duration is important for hedging. For interest rate dynamic, we assumed the Bachelier model.

$$dr(t) = \alpha * \Delta t + \sigma\sqrt{\Delta t}W \tag{12}$$

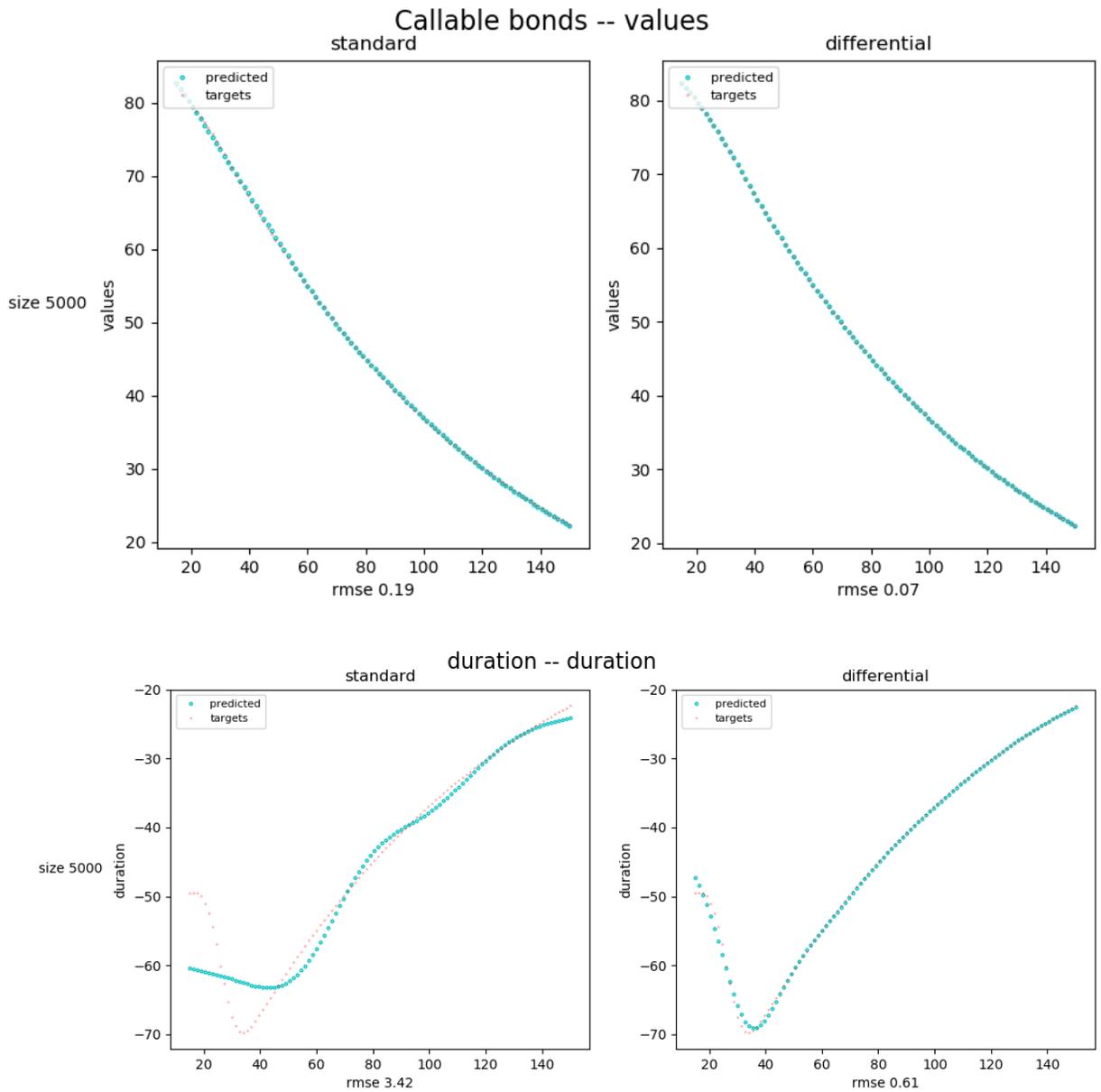

Differential neutral network catches dynamic of duration more accurately.

5. We validated the approach for large portfolio like 250, 500 and 1000 instruments. Directly to learn 'greeks' for large portfolios is highly expensive task from computational point of view. We tried to validate methodology of Differential PCA. Differential PCA remove irrelevant factors and considerably reduce dimension. As a data preparation step, differential PCA may significantly reduce dimension, enabling faster, more reliable training of neural networks. Differential PCA is a useful algorithm on its own right, providing a low dimensional latent representation of data on orthogonal axes of relevance. In our case it dramatically shrunk dimensions, it is known in the case correlated Bachelier model [4], we just tried to train differential neural network in this case. We learned differential neural network in reduced space. We trained networks on 100K paths for portfolio with 250, 500, 1000 instruments.

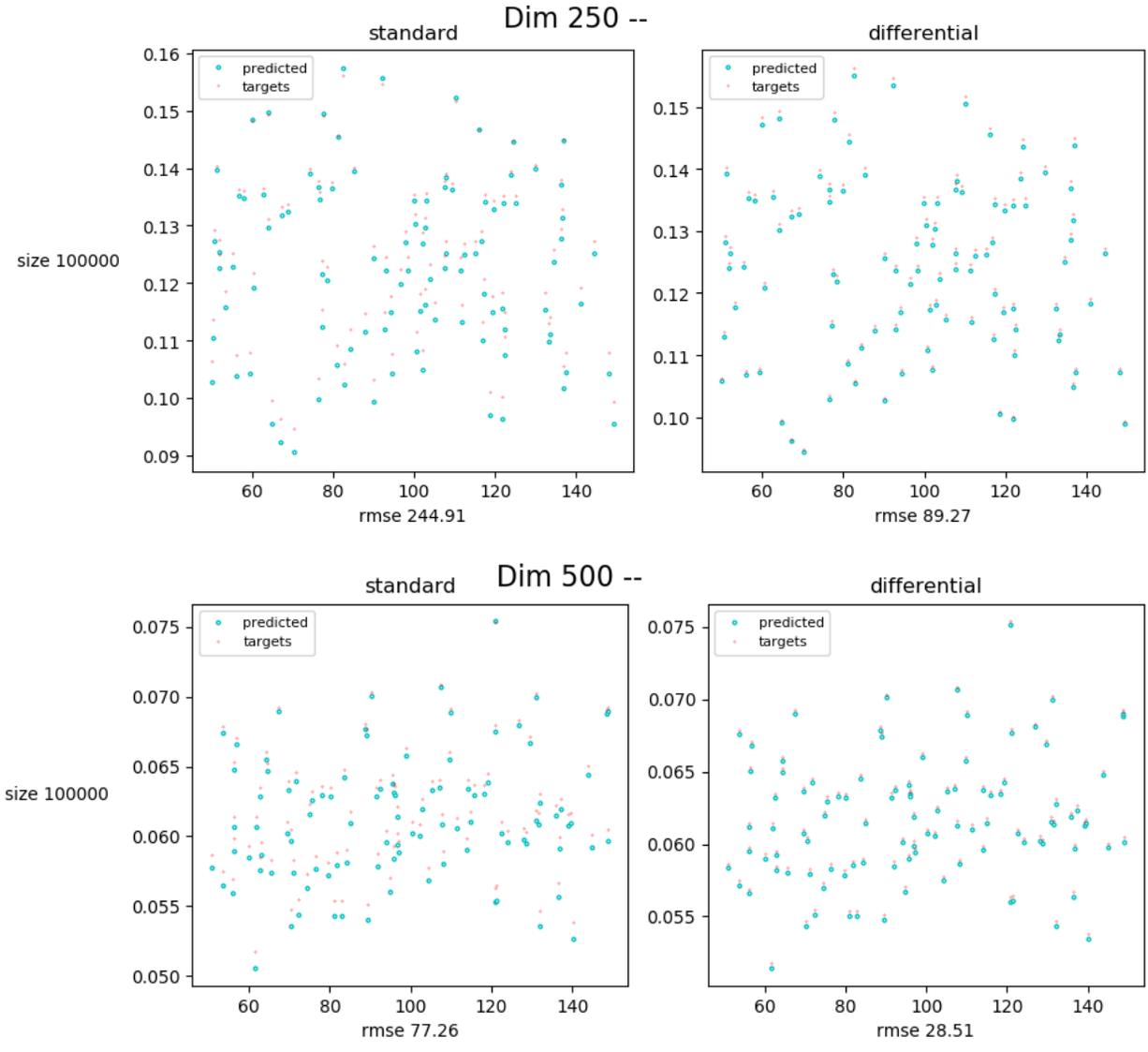

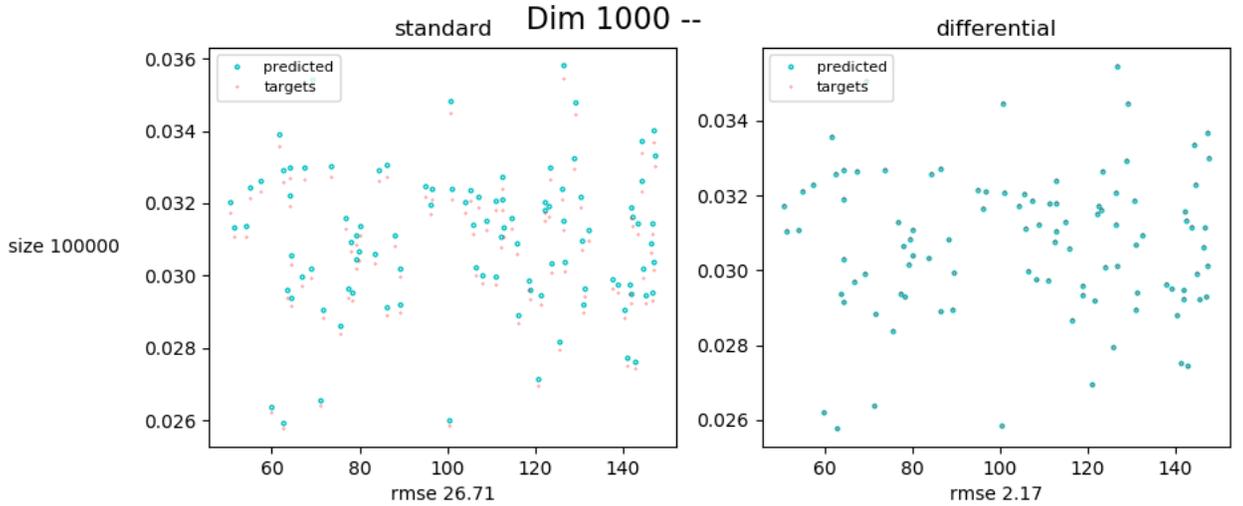

In all cases accuracy of 'greeks' for differential machine learning is higher than feed forward net.

6.  We considered Libor Market Model by Prof Mike Giles, Mathematical Institute, University of Oxford. The Libor Market Model (LMM), also known as BGM, is a widely used interest rate term structure model. As an extension of the Heath, Jarrow, and Morton (HJM) model on continuous forward rates, the LMM takes market observables as direct inputs to the model. Whereas the HJM model describes the behavior of instantaneous forward rates expressed with continuous compounding. LMM postulates dynamical propagation of the forward Libor rates, which are the floating rates to index the interest rate swap funding legs. LIBOR market model uses forward LIBOR rates as fundamental assets. Let $L_t^i$ denote the forward LIBOR rate over the time interval $[T_i, T_{i+1}]$, where $T_i, i = 0,1, \dots, N$, are LIBOR reset dates. Each forward LIBOR rate has the following dynamics [2].

$$dL_t^i = L_t^i \left( \mu_t^i \; dt + \; \sigma_t^i \; dW_t^i \right) \qquad (13)$$

where $W_t^i$ is the Brownian motion for $L_t^i$ and forward LIBOR rates are allowed to have factor correlations, $dW_t^i \, W_t^j = \rho_t^{ij}, \forall i,j$. We assume constant volatilities, $\sigma_t^i = \sigma$. We also assume all LIBOR rates share the same Brownian factor $W_t$. The discretized version of forward LIBOR rate then reads

$$L_{T_{j+1}}^i = L_{T_j}^i \, exp\left[ \sigma^i \left( \sum_{k=j+1}^{i} \frac{\tau_k \, L_{T_j}^k \, \sigma^k}{1 + \tau_k \, L_{T_j}^k} - \frac{1}{2}\sigma^i \right) \tau_j + \sigma^i \sqrt{\tau_j} \, Z_j \right] \qquad (14)$$

With equation (14) we can simulate forward LIBOR rates at particular time points and price options whose payoffs depend on LIBOR rates. We build TensorFlow graph of calculation for each scenario in Monte Carlo simulation, the dependence of payoff from initial values of LIBOR forward rates. In this case we consider the case of Caplet payoffs, but it does not matter, we can easily change it to another payoff. Because we build graph of calculation for LIBOR model, then we get gradients for LIBOR model and construct train set to learn differential neutral network. In this case differential Neutral Network outperforms feed forward network. We estimate Caplets values (payoffs). [2] There are values of them.

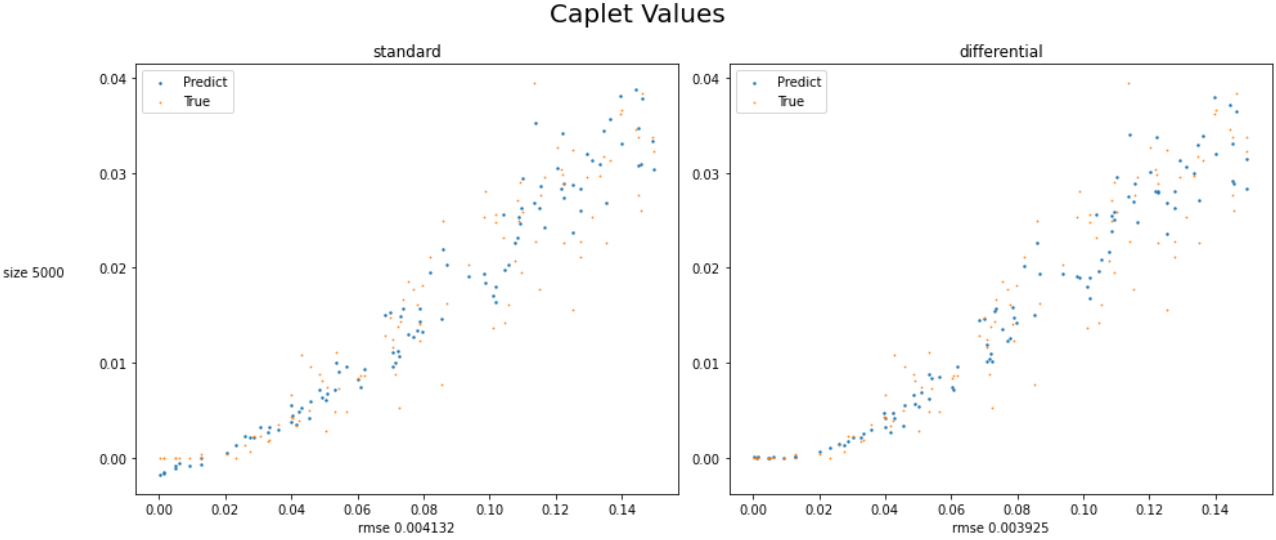

There are derivatives of Caplet payoffs.

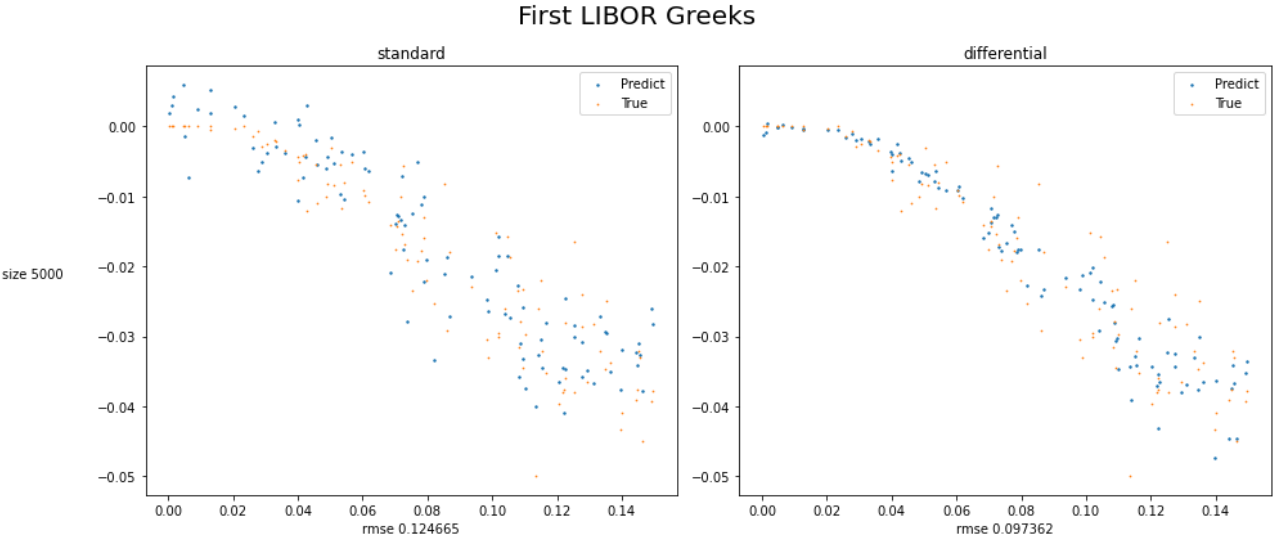

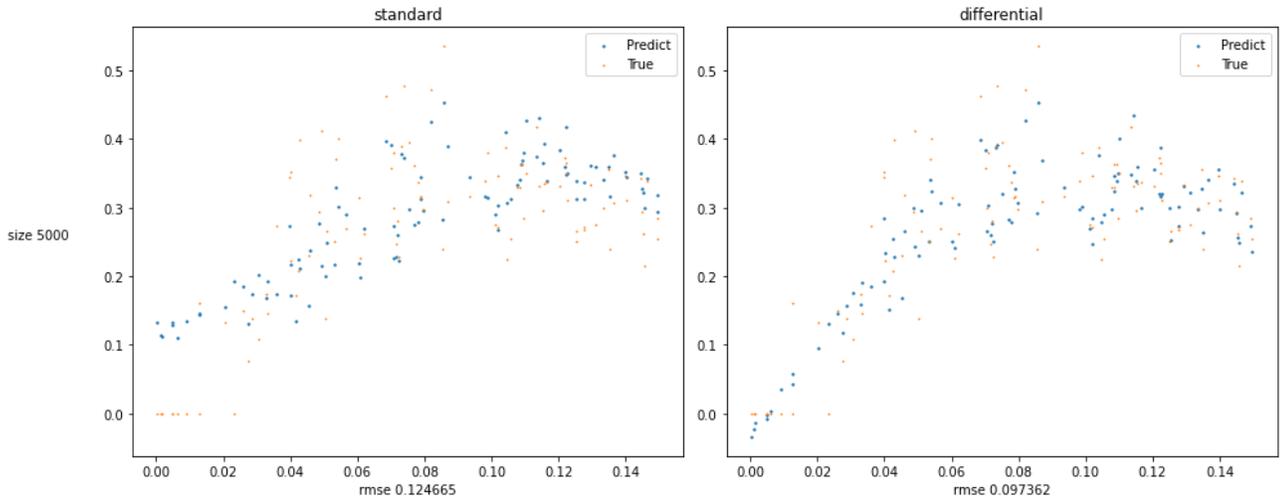

Let's compare them in terms of loss functions. The neural network has hidden size 64 and the number of hidden layers 6 for both standard and differential neural networks.

Caplet payoffs from the training dataset were generated with 10 Monte Carlo paths, from the test dataset with 10000 Monte Carlo paths.

Experiments were carried out for different strikes, different time to maturity (TTM) and different amounts of forward LIBOR rates. Some examples of the learning curves with multiple restarts on test samples are shown in following graphs.

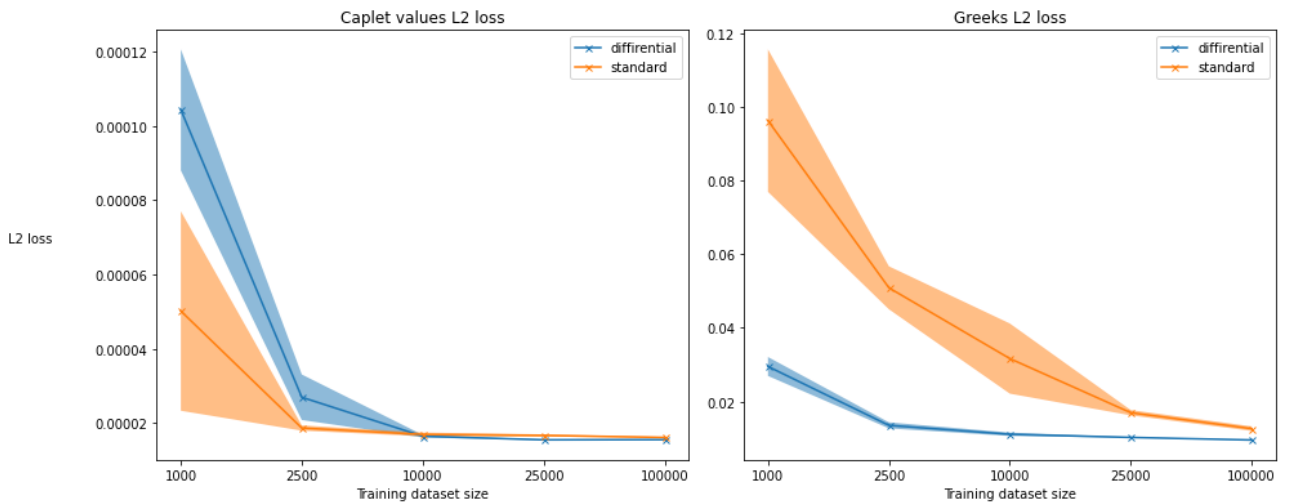

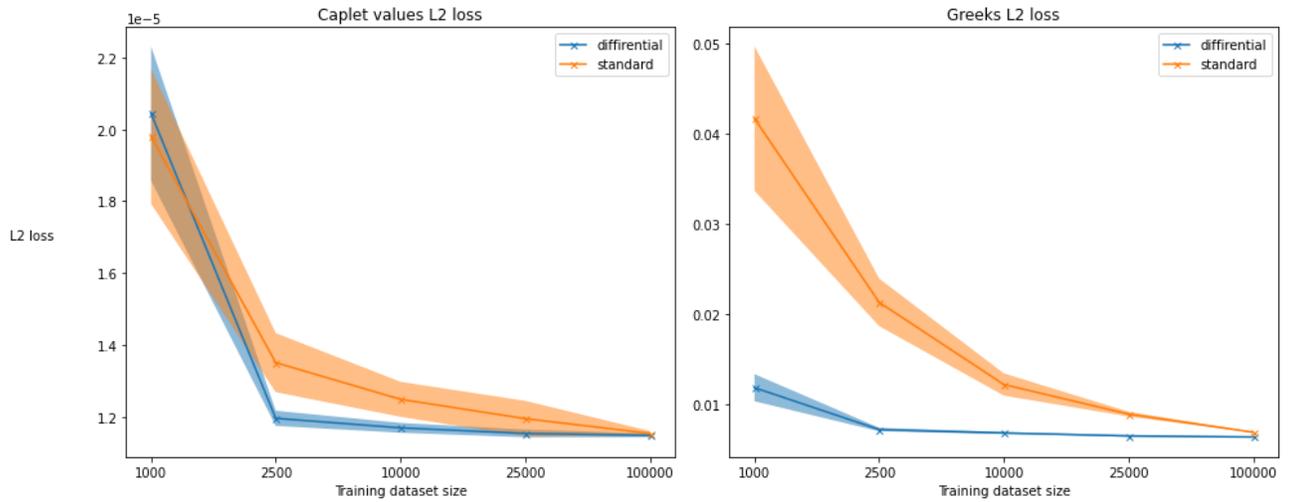

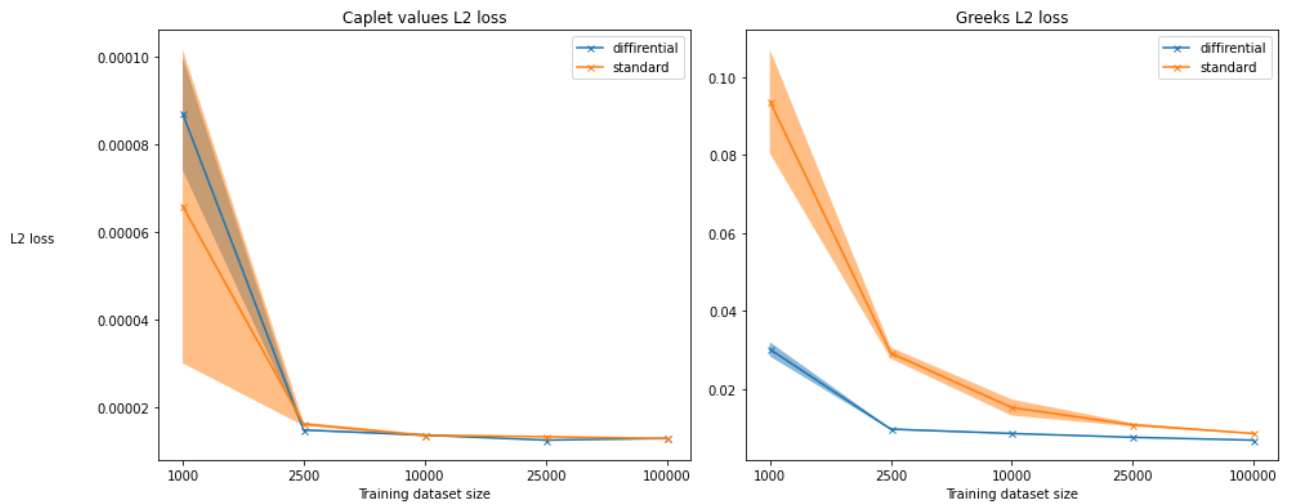

General conclusion for all graphs: the differential network has a big advantage in terms of Greeks L2 loss on a small data set and has a slight disadvantage in terms of Caplet payoff L2 loss as compared to a standard network. On large data set sizes, the differential network always has an advantage in terms of Greeks L2 loss (significant or not) and has a comparable quality in terms of Caplet payoff L2 loss compared to a standard network.

Moreover, quite often the differential network has an advantage in the quality of the predicted Caplet payoffs.

It can be assumed that additional information about the gradients helps to give more information to the approximator about the behavior of the function in the neighborhood of points, and this gives an increase in quality even though the weights of this network have to share information both about Caplet payoffs and derivatives.

Let's test this hypothesis by increasing the hidden size to 128 and the number of hidden layers to 10 for both networks.

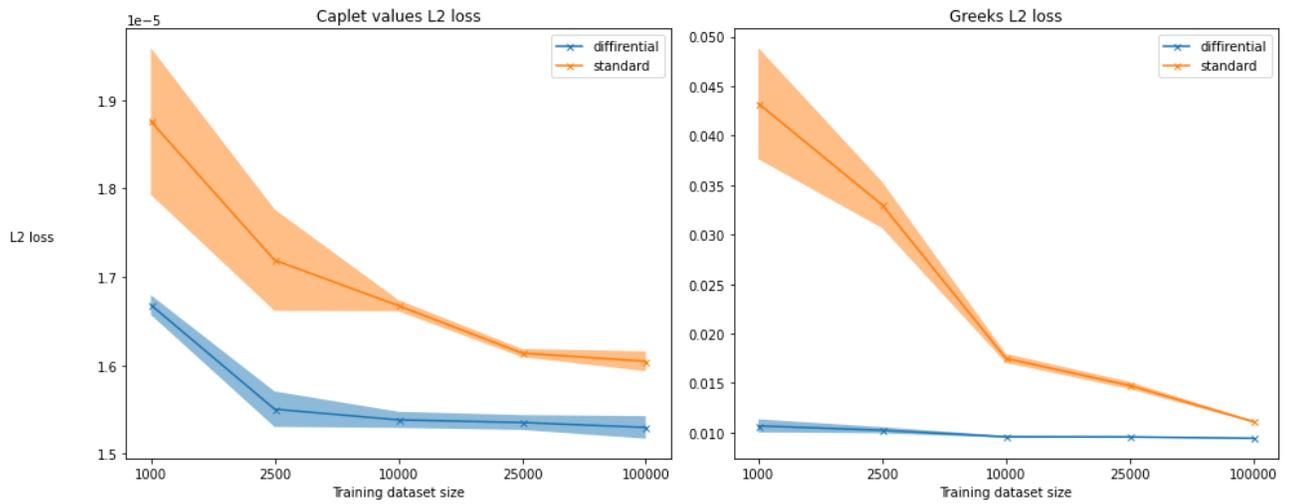

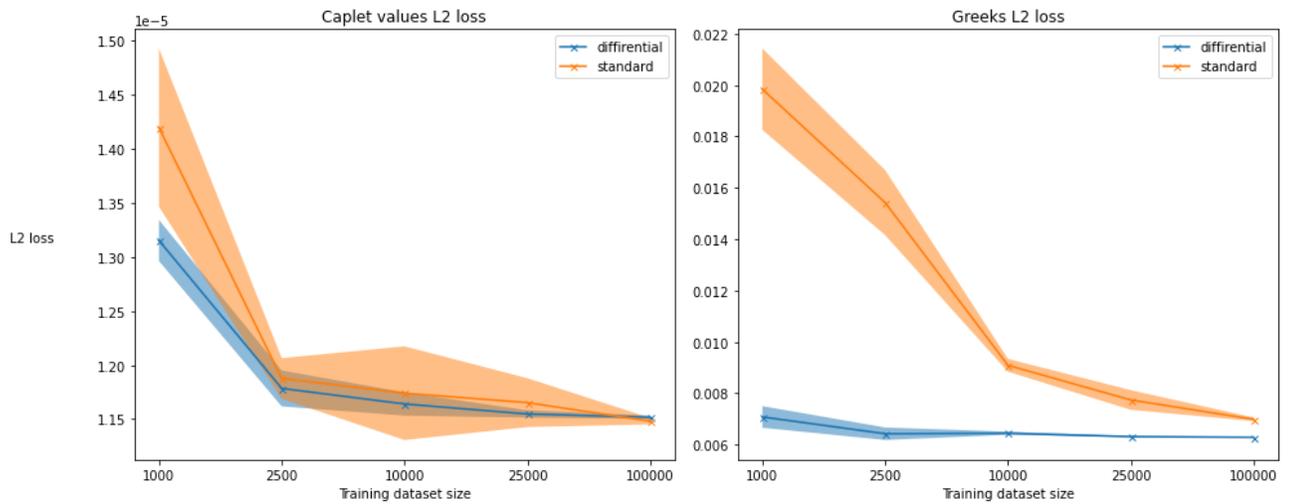

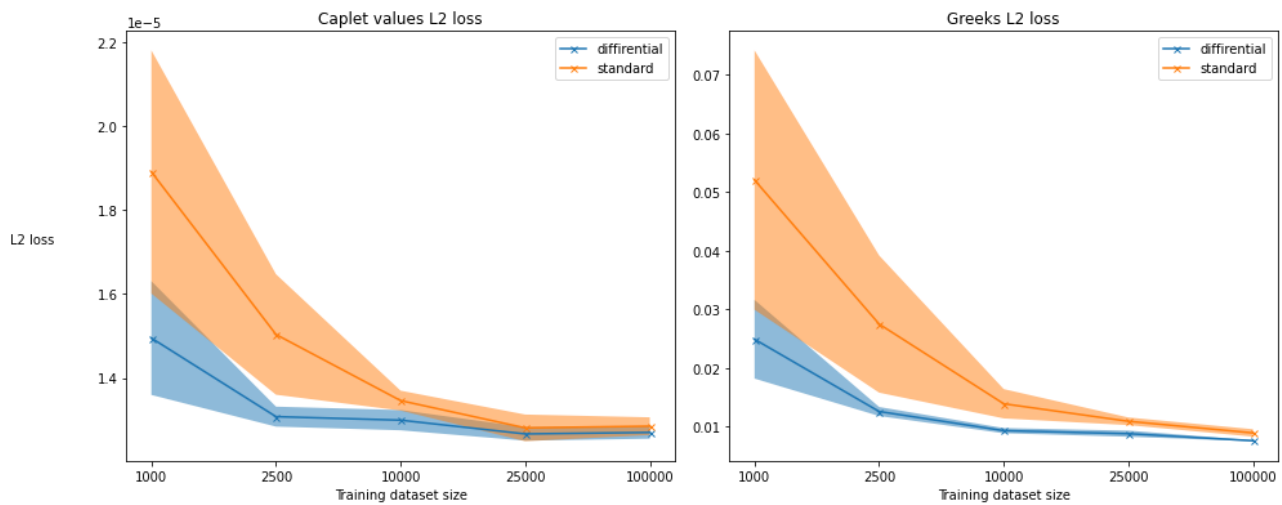

In this case, the differential network shows a significant advantage relative to the nondifferential network, so we can make sure that the derivatives help to get more information about the behavior of the function, and this can be used to train the neural network.

It can also be concluded that the use of derivatives differs from the use of additional data not only that the quality of the predicted derivatives is much higher, but also that in order to get an advantage on the values of the Caplet payoffs, it may be necessary to increase the neural network. The reason for this: it is necessary to approximate two loss functions at once.

7. Also, we dealt with SABR model. It is the model of stochastic volatility. The SABR model helps explain the volatility smile better and resolve the problem of unstable hedges.

$$\begin{cases} dS = \sigma dW_1 \\ d\sigma = \alpha dW_2 \\ <dW_1 dW_2> = \sigma \end{cases} \quad (15)$$

There are options values.

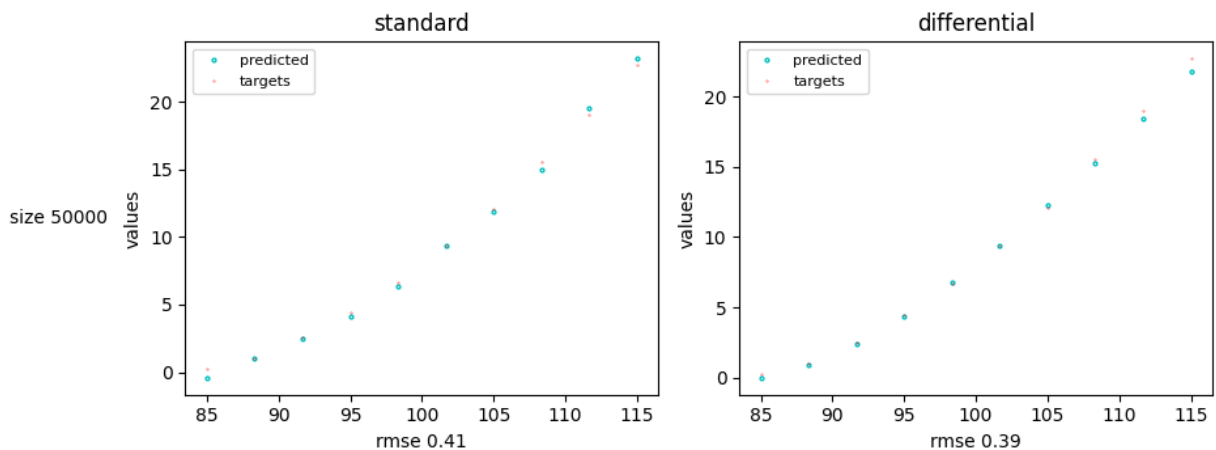

There are greeks.

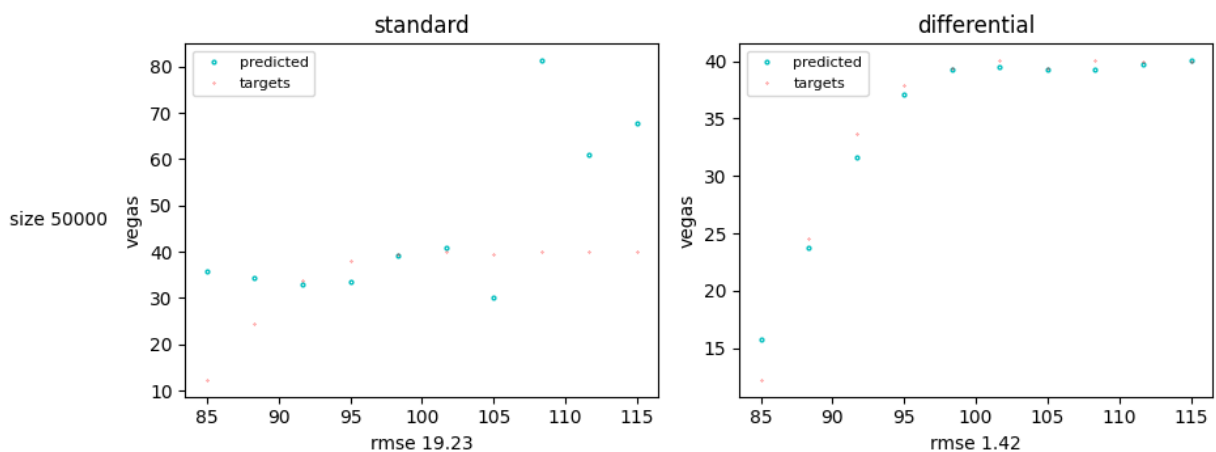

The accuracy of differential neural network is higher again.

8. Also, we considered worst-of options. We estimated worst-of options for basket of n correlated stocks. Prices are again generated by means follow stochastic differential equation (10). It is case of 5 stocks. Differential neural network works better than feed forward network.

Worst-of options --

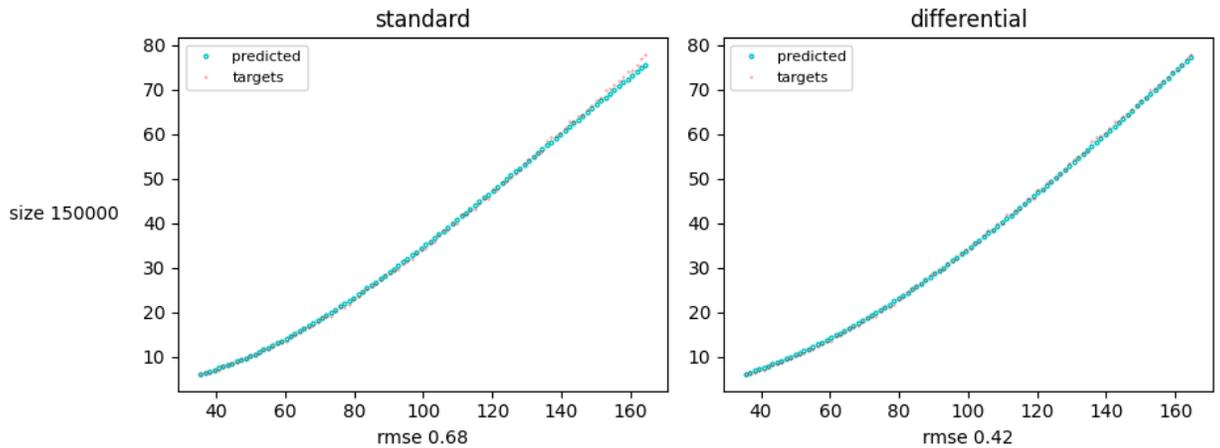

deltas --

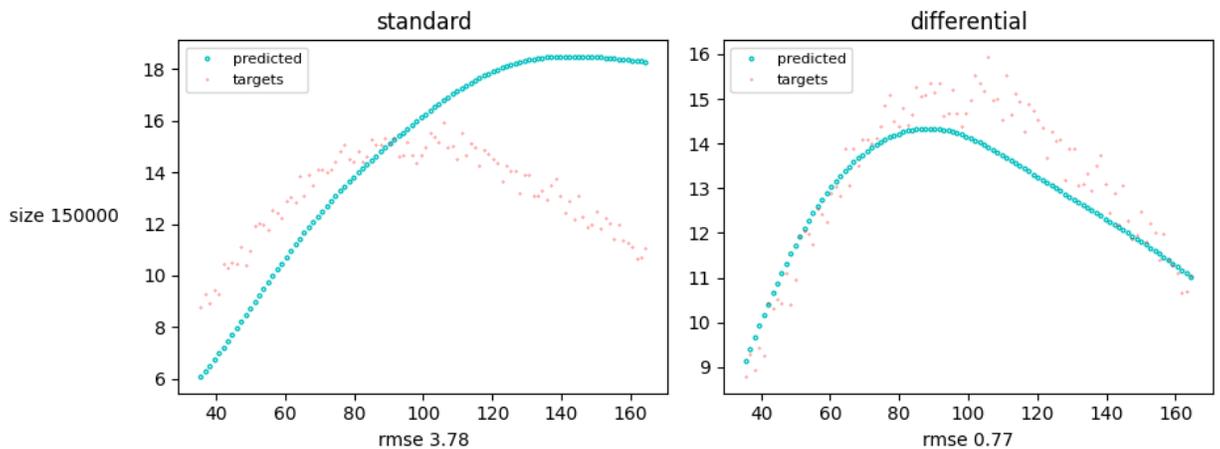

## Conclusion

In all the cases selected by us to test the approach [4], differential neural network provided much better accuracy and convergence rate than feed forward network.

It was shown that sensitivities of Asian option in the case of volatility curve are estimated with good accuracy.  It was empirically tested that the higher the order of volatility, the stronger advantage of differential neural network.

We also estimated second derivative of payoff by mean likelihood method [5]. Fast and accurate estimate of gamma is important for gamma trading.

Page | 17

We also successfully applied the approach to instrument portfolios. We reduced dimension of input risk factors via differential PCA [4].

We considered Libor Market Model by Prof Mike Giles [2]. We built TensorFlow calculation graph of one Monte Carlo path for forward LIBOR rates and got gradients and used it as supervisor. It is new example how simply we can resolve very complicated task by means TensorFlow and differential machine learning. The approach could be applicable to this can be useful for any interest rate derivatives [2].

We considered the case of worst-of options. It is very popular option and it is very important accurately pricing this option and greeks.